\documentclass[11pt]{article}
\usepackage[a4paper,margin=1in]{geometry}
\usepackage[T1]{fontenc}
\usepackage[utf8]{inputenc}
\usepackage{lmodern}
\usepackage{amsmath,amsfonts,amssymb,amsthm}
\usepackage{graphicx}
\usepackage{xcolor}
\usepackage{hyperref}
\usepackage{tikz}
\usetikzlibrary{positioning}

\hypersetup{
  colorlinks=true,
  linkcolor=black,
  citecolor=black,
  urlcolor=blue
}
\title{Born's Rule from Reversible Evolution and Record Formation}
\author{Oskar Axelsson\\
Independent Researcher, H\"o\"or, Sweden\\
\texttt{oskar@oascitech.se}\\
ORCID: 0000-0003-3209-7241}
\date{}

\begin{document}
\maketitle

\begin{center}
\small Published in \emph{Journal of Modern Physics} \textbf{17}(10), 1282--1294 (2026).\\
\href{https://doi.org/10.4236/jmp.2026.1710056}{doi:10.4236/jmp.2026.1710056}
\end{center}

\begin{abstract}
We show that the quadratic measure associated with the Born rule need
not be introduced as an independent probabilistic postulate, but follows,
under the stated structural and regularity assumptions, from the
compatibility between reversible linear evolution and the formation of
stable records.

Physical processes alternate between reversible dynamics and episodes
of record formation. Quantities that remain well defined throughout
such processes must therefore admit a representation that remains
consistent across both regimes. Prior to record formation, alternatives
combine additively at the level of complex amplitudes, whereas
successive record-forming distinctions admit a multiplicative
representation through refinement.

Requiring compatibility between these compositional structures
constrains the admissible weight assignment to the family
$|\alpha|^p$. The finite-dimensional specialization of Lamperti's
isometry theorem then singles out $p=2$ as the unique value compatible
with continuous reversible linear mixing while preserving total
weight.

The Born rule therefore emerges as the unique weight assignment
compatible with reversible linear composition, multiplicative record
refinement, phase invariance, and continuous reversible invariance,
without invoking probabilistic assumptions or a specific interpretation
of quantum measurement.
\end{abstract}

\bigskip
\section{Introduction}

The Born rule plays a central role in quantum theory, connecting the
mathematical formalism of amplitudes with experimentally observed
outcome frequencies. In the standard formulation it is introduced as a
postulate, and considerable effort has been devoted to deriving it from
more primitive physical principles.

Several approaches have been proposed. Decision-theoretic arguments in
the Everett interpretation attempt to recover the rule from rational
consistency requirements \cite{Deutsch1999}. Envariance-based
derivations relate the rule to symmetry properties of entangled states
\cite{Zurek2005}. Other approaches derive the quadratic measure from
structural features of Hilbert space or Gleason-type theorems
\cite{Gleason1957}. Derivations based on time-reversal symmetry have also been proposed \cite{Ilyin2016}. The present approach differs in deriving a general weight assignment from the compatibility between reversible evolution and irreversible record formation, without assuming a probabilistic interpretation from the outset.

As emphasized in several reviews, derivations of the Born rule
generally require assumptions beyond the standard dynamical postulates
of quantum mechanics \cite{Vaidman2020}. At the same time, the role of
these assumptions remains a matter of debate. Recurring criticisms
include the implicit introduction of probabilistic concepts through
auxiliary assumptions and the absence of a clear account of physical
outcomes within the formalism. A recent review likewise highlights the
variety of existing formulations of the Born rule, their differing
domains of validity, and the continuing lack of a universally accepted
account of the emergence of definite physical outcomes
\cite{Neumaier2025}.

The present derivation differs from representative existing approaches
primarily in the assumptions from which it begins. It does not invoke
decision-theoretic rationality, probabilistic symmetry arguments,
expectation values, or a pre-existing quadratic measure. Instead, it
starts from reversible linear composition, persistent record formation,
and consistency of the weight assignment across these two regimes. A
broader comparison of existing Born-rule derivations and their
assumptions is provided in Refs.~\cite{Vaidman2020,Neumaier2025}.

The present work addresses these concerns from a different
perspective. Probabilities are not introduced as primitive quantities,
and outcomes are represented by stable records, understood as persistent physical outcomes that can be identified after their formation. Recent analyses have also emphasized the
indispensable role of the superposition principle, that is, the additive composition of amplitudes, in many proposed derivations of the
Born rule \cite{Zhang2026}. In the present work, additive composition
is not derived independently but taken directly from the linear
structure of reversible quantum evolution.

We therefore pursue a structural route based on the compatibility
between reversible evolution and irreversible record formation.
Reversible evolution is described by the Schrödinger equation and
possesses continuous symmetries together with the associated
conservation laws. Physical processes, however, are not composed
exclusively of reversible evolution, but repeatedly alternate between
reversible dynamics and episodes of record formation.

Both ingredients are independently motivated by established physical
practice. Reversible evolution is a central element of the standard
quantum formalism, while stable records are the operational basis by
which outcomes are identified and compared in experiment. The physical starting points of the present derivation are therefore structures that already play a fundamental role in the description of physical processes, rather than
on additional assumptions concerning observers, decision-making, or
subjective probabilities. We do not attempt to explain the origin of irreversibility or the arrow of time. Rather, we start from the empirical coexistence of reversible quantum evolution and persistent records, and investigate the constraints imposed by requiring that quantities associated with possible outcomes remain consistently represented across both regimes.

If quantities such as energy remain well defined throughout a physical
process, their representation cannot be determined independently in
these two regimes. Reversible and irreversible processes must therefore
admit a common structural description. The present work leverages the
constraints imposed by this requirement on admissible outcome weights.

In this setting, reversible evolution combines alternatives additively
at the amplitude level, whereas record formation composes
multiplicatively through successive refinement, whereby each new record further specifies the realized outcome. Any
quantity that remains consistently represented across both regimes must
therefore be compatible with these distinct compositional structures.

The distinction between reversible evolution and irreversible record-forming processes has been explored in various forms in the
literature. In particular, related ideas appear in the work of Horodecki, where reversible and irreversible aspects of quantum
processes are discussed in a broader informational context \cite{Horodecki2025}.

The derivation presented here is formulated within the standard
framework of quantum mechanics, assuming reversible linear evolution with complex amplitudes together with the occurrence of irreversible record formation. No specific interpretation or mechanism for this process is
assumed. Nor does the derivation assume probabilities or expectation values as
primitive concepts. Instead, the admissible weight assignment is derived
from the compatibility of reversible evolution and irreversible record
formation. A probabilistic interpretation is considered only after the quadratic form has been derived.

We show that additive reversible composition and multiplicative record
refinement, together with the requirement that outcome weights remain
consistently represented across both regimes, uniquely select a
quadratic assignment of outcome weights. The Born rule therefore
emerges as the unique weighting compatible with reversible linear
evolution and multiplicative irreversible refinement. To make this argument precise, we first identify the structural
features of the reversible and irreversible regimes separately before
examining the constraints imposed by their compatibility.

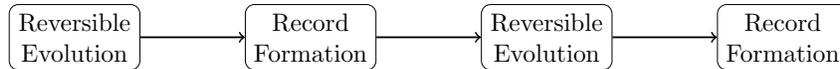
\begin{figure}[t]
\centering
\resizebox{0.9\linewidth}{!}{
\begin{tikzpicture}[
node distance=2.5cm,
box/.style={draw, rectangle, rounded corners, align=center, minimum width=2.0cm, minimum height=0.9cm},
arrow/.style={->, thick}
]

\node[box] (rev1) {Reversible\\Evolution};
\node[box, right=1.6cm of rev1] (rec1) {Record\\Formation};
\node[box, right=1.6cm of rec1] (rev2) {Reversible\\Evolution};
\node[box, right=1.6cm of rev2] (rec2) {Record\\Formation};

\draw[arrow] (rev1) -> (rec1);
\draw[arrow] (rec1) -> (rev2);
\draw[arrow] (rev2) -> (rec2);

\end{tikzpicture}
}
\caption{
Physical evolution alternates between reversible dynamics and the
formation of stable records. Reversible evolution is linear and time symmetric, while record formation
produces effectively irreversible outcomes. Each record distinguishes a
single realized outcome, after which reversible evolution continues
from the recorded configuration.
}
\label{fig:record_sequence}
\end{figure}

\section{Two Operational Regimes}

Measurement interactions are commonly described as involving two
operationally distinct regimes: reversible linear evolution prior to
record formation and effectively irreversible formation of persistent
records associated with measurement outcomes
\cite{Schlosshauer2005}. This operational distinction is widely used
in both textbook and research treatments of quantum measurement.

\begin{enumerate}
\item \textbf{Reversible regime.}
Prior to formation of a stable record, evolution is reversible and
linear at the amplitude level. States associated with different
potential outcomes remain operationally interconvertible.

\item \textbf{Irreversible regime.}
Once a stable record forms, configurations corresponding to distinct
outcomes become operationally distinguishable and are no longer
reversibly interconvertible.
\end{enumerate}

Here irreversibility refers to operational irreversibility: the
recorded configurations cannot be returned to their pre-record state
by physically accessible reversible transformations, even though the
underlying microscopic dynamics may remain time symmetric.

For clarity, the derivation below uses the following physical (\textbf{P}) and
mathematical \textbf{(M}) assumptions.

\begin{enumerate}
\item[\textbf{P1.}] \textbf{Reversible linear composition.}
Prior to record formation, alternatives are represented by complex
amplitudes and combine linearly.

\item[\textbf{P2.}] \textbf{Persistent record formation.}
Physical processes may produce stable records that render distinct
outcomes operationally distinguishable and whose recorded distinction remains available under subsequent evolution and is not erased by the reversible transformations under consideration.

\item[\textbf{P3.}] \textbf{Cross-regime consistency.}
A non-negative weight assigned to a possible record before its
formation retains the same physical meaning when that record becomes
established.

\item[\textbf{P4.}] \textbf{Phase invariance.}
States differing only by an overall complex phase receive the same
weight.

\item[\textbf{M1.}] \textbf{Regularity.}
The weight assignment and the composition laws introduced below are
taken to be continuous on their stated domains. Where stronger
regularity properties are required, they are stated explicitly at
the point of use.
\end{enumerate}

No probabilistic interpretation of the weight is assumed at this
stage. Additional structural requirements associated with sequential
record refinement and reversible invariance are introduced explicitly
where they enter the derivation.

\subsection{Reversible Composition}

At the microscopic level, the dynamical laws governing physical
systems are reversible. If a quantum state evolves according to a deterministic dynamical law, the same law also determines how the earlier state can be reconstructed from its later state.

A familiar example is the Schrödinger equation,
\begin{equation}
i\hbar \frac{\partial \psi}{\partial t} = H \psi ,
\end{equation}

which generates unitary evolution.
Given a solution $\psi(t)$, the state at an earlier time can be
obtained by evolving with the inverse unitary operator
$U^{-1}(t)=U(-t)$. The equation itself therefore contains no intrinsic
direction of time.

This microscopic reversibility contrasts with everyday experience. Macroscopic processes such as measurement or record formation appear
irreversible because they involve the creation of persistent records that are impossible to erase in practice through local physical
interactions. Such irreversibility reflects operational limitations rather than a fundamental asymmetry of the underlying dynamical laws.

The coexistence of reversible evolution and irreversible record
formation raises the question of how quantities that remain well
defined throughout a physical process can be represented consistently
across both regimes. This compatibility requirement imposes strong constraints on how non-negative weights, rather than probabilities, may be assigned to possible outcomes. The term weight is used because no probabilistic interpretation is assumed a priori; the relation to probability emerges only after the quadratic form has been established.

The present work does not attempt to derive the additive composition of amplitudes, which is taken directly from the linear structure of reversible quantum evolution \cite{Zhang2026}.

If two alternatives lead to the same subsequent state without leaving
a record of their distinction, they cannot be operationally
distinguished and must be represented by a single effective
alternative. Since reversible evolution is linear, the corresponding
amplitudes combine additively,

\begin{equation}
\alpha_{\mathrm{combined}} = \alpha_1 + \alpha_2.
\end{equation}

Addition is therefore the fundamental composition law of the reversible
regime.

Reversible evolution admits a continuous symmetry corresponding to
global phase transformations,
\begin{equation}
\alpha \mapsto e^{i\theta}\alpha .
\end{equation}

A weight is a non-negative quantity assigned to a possible persistent
record prior to its realization. The realized record retains this same
assignment upon record formation. Throughout the present work, the term
\emph{weight} refers to this single assignment both before and after
record formation.

States differing only by a global phase are operationally indistinguishable prior to record formation and must therefore be assigned the same weight. Outcome weights are therefore invariant under global phase transformations.

It follows that weights cannot depend on the overall phase of
$\alpha$, and can therefore be written as a function of its magnitude,

\begin{equation}
\label{mu_magnitude}
\mu = f(|\alpha|).
\end{equation}

\subsection{Irreversible Regime}

The preceding discussion identifies the composition law and symmetry
properties of the reversible regime. We now turn to the corresponding
structure associated with irreversible record formation.

Physical processes consist of alternating stages of reversible evolution
and irreversible record formation. At discrete stages, irreversible record formation occurs, selecting
a single outcome from a set of alternatives. No specific mechanism for this selection is assumed.

Once a stable record forms, configurations become operationally distinguishable. Subsequent evolution proceeds
within the configuration consistent with the recorded outcome until a further irreversible record forms.
Upon record formation, the realized record inherits the corresponding weight $\mu$.

The present derivation does not assume a particular microscopic mechanism
by which persistent records are established. In modern quantum theory,
effective record formation is commonly described in decoherence-based
approaches through environmental interaction, where the global evolution
remains reversible while local subsystems acquire effectively irreversible
records \cite{Zurek2003,Zurek2005}. The present argument is compatible
with such descriptions. It requires only that a regime exists in which
persistent records can be treated as operationally established. The
derivation therefore concerns the structural consequences of record
formation rather than its microscopic origin.

\subsection{Consistency of Sequential Refinement}

Successive record formation progressively specifies the realized
outcome: each new record introduces information not contained in
previous records. When multiple record-formation events occur in
sequence, the final record encodes all distinctions introduced at
each stage.

Let $\mu(R)$ denote the non-negative weight assigned to a record $R$.
We require that this assignment respect the compositional structure
of successive refinement. In particular, the weight assigned to a
final record must not depend on how the refinement leading to that
record is partitioned into intermediate stages.

Suppose the composition law for successive refinements is represented
by an operation $\circ$,

\begin{equation}
\mu(R_{12})
=
\mu(R_1)\circ\mu(R_2).
\end{equation}

For positive weights, we require the composition law $\circ$ to be
continuous, associative, and strictly increasing in each argument.
Associativity expresses independence of the grouping of successive
refinements,

\begin{equation}
(\mu_1\circ\mu_2)\circ\mu_3
=
\mu_1\circ(\mu_2\circ\mu_3),
\end{equation}

while strict monotonicity expresses that increasing the contribution
of either refinement increases the weight of the cumulative record.

\paragraph{Proposition.}
Let $\circ$ be a binary operation on a real interval of positive
weights. If $\circ$ is continuous, associative, and strictly increasing
in each argument, then, up to a continuous strictly monotonic
reparameterization, it admits an additive representation.

By Aczél's representation theorem for continuous, associative,
strictly monotonic operations on a real interval
\cite{Aczel1948}, there exists a continuous strictly monotonic
function $h$ such that

\begin{equation}
h(\mu_1\circ\mu_2)
=
h(\mu_1)+h(\mu_2).
\end{equation}

Defining the reparameterized weight

\begin{equation}
\widetilde{\mu}
=
\exp[h(\mu)],
\end{equation}

therefore gives

\begin{equation}
\widetilde{\mu}(R_{12})
=
\widetilde{\mu}(R_1)\widetilde{\mu}(R_2).
\end{equation}

We henceforth choose this multiplicative representation and suppress
the tilde, writing

\begin{equation}
\mu(R_{12})
=
\mu(R_1)\mu(R_2).
\end{equation}

The distinguished zero-weight case is treated separately as the
boundary corresponding to absence of an alternative. Multiplication
extends continuously to this boundary through

\begin{equation}
0\cdot\mu=\mu\cdot0=0.
\end{equation}

Thus multiplication is not assumed as the primitive refinement law.
It is the canonical representation, up to continuous monotonic
reparameterization, of any continuous, associative, strictly
monotonic composition law for positive refinement weights.

This relation expresses the cumulative character of record formation:
each stage contributes an additional distinction while preserving
those already encoded in previous records.

\section{Compatibility Between Regimes}

We now examine how the two regimes must connect. The central
question is how quantities that remain well defined throughout a
physical process can survive transitions between reversible
evolution and record formation.

In the reversible regime, alternatives combine additively at the
amplitude level. In the irreversible regime, distinctions accumulate
through multiplicative refinement. The compatibility problem is
therefore to identify a weight assignment that remains consistently
defined across regimes whose fundamental composition laws are
addition and multiplication, respectively.

The refinement associated with record formation changes the relative
prominence of an existing alternative without altering its identity.
For a nonzero amplitude magnitude, we characterize a particular
refinement by its relative change

\begin{equation}
\lambda
=
\frac{|\alpha|' }{|\alpha|},
\qquad \lambda>0,
\end{equation}

so that, equivalently,

\begin{equation}
|\alpha|'=\lambda|\alpha|.
\end{equation}

This relation defines the refinement factor $\lambda$; it does not
assert that every continuous transformation fixing zero is intrinsically
linear.

Successive refinement factors compose multiplicatively. If

\begin{equation}
|\alpha|'=\lambda_1|\alpha|,
\qquad
|\alpha|''=\lambda_2|\alpha|',
\end{equation}

then

\begin{equation}
|\alpha|''
=
\lambda_2\lambda_1|\alpha|.
\end{equation}

To represent the same refinement consistently at different initial
amplitude magnitudes, we further require that its effect on the weight
depend only on the refinement factor $\lambda$, rather than separately
on the initial magnitude. Thus

\begin{equation}
\label{eq:fLambdaAlpha}
f(\lambda|\alpha|)
=
g(\lambda)f(|\alpha|).
\end{equation}

Successive rescalings do not represent independent physical dynamics.
Rather, they provide a mathematical representation of successive
stages of record formation. Each stage further suppresses competing
alternatives while preserving the realized alternative. If the
description is to remain independent of how the refinement is
partitioned into intermediate stages, successive rescalings must
compose consistently in the same manner as the corresponding sequence
of record refinements.

Since weights represent physically meaningful quantities associated
with continuously varying amplitudes, we assume that the scaling
function $g(\lambda)$ is positive and satisfies the mild regularity
condition of continuity (or, equivalently for the present purpose,
measurability). These assumptions exclude pathological solutions
without restricting physically admissible behavior.

Consistency therefore requires

\begin{equation}
g(\lambda_1\lambda_2)
=
g(\lambda_1)g(\lambda_2).
\end{equation}

This is the standard multiplicative functional equation, whose
positive continuous (or measurable) solutions are

\begin{equation}
g(\lambda)=\lambda^p,
\end{equation}

for some real $p>0$.

Substituting this result into Eq. \ref{eq:fLambdaAlpha} gives

\begin{equation}
f(\lambda |\alpha|)
=
\lambda^p f(|\alpha|).
\end{equation}

Since this relation holds for every positive scaling parameter
$\lambda$, for $|\alpha|>0$ we may choose

\begin{equation}
\lambda=\frac{1}{|\alpha|}.
\end{equation}

This choice specifies the scaling transformation; it does not impose
a condition on the original amplitude. Under this particular
transformation,

\begin{equation}
|\alpha|'=\lambda|\alpha|=1.
\end{equation}

Consequently,

\begin{equation}
f(1)=|\alpha|^{-p}f(|\alpha|),
\end{equation}

or equivalently,

\begin{equation}
f(|\alpha|)=f(1)|\alpha|^p.
\end{equation}

Absorbing the constant factor $f(1)$ into the normalization yields

\begin{equation}
\mu = f(|\alpha|)
=
|\alpha|^p.
\end{equation}

Compatibility between reversible evolution and record formation
therefore restricts the admissible weight assignments to the family
$\mu = |\alpha|^p$. To determine which member of this family is
physically admissible, we must now impose the additional requirement

that continuous reversible evolution preserve the total weight
assigned to mutually exclusive alternatives.

\section{Reversible Invariance}

Consider a finite set of $n$ mutually exclusive alternatives, represented
by the complex amplitude vector

\begin{equation}
\boldsymbol{\alpha}
=
(\alpha_1,\ldots,\alpha_n)
\in \mathbb{C}^n.
\end{equation}

Prior to record formation, states evolve according to reversible linear
dynamics. Such transformations are represented by invertible
complex-linear maps

\begin{equation}
\boldsymbol{\alpha}'
=
U\boldsymbol{\alpha},
\end{equation}

where $U$ is an invertible $n\times n$ complex matrix.

By continuous reversible evolution we mean continuous families of such
invertible linear transformations, including transformations that
smoothly mix distinct amplitude components.

The weight assignment must be compatible with this reversible
evolution.

A reversible transformation changes the amplitudes describing a state but does not itself create or remove possible outcomes. We therefore require that, if the same weight assignment is to remain consistently defined during reversible evolution, reversible transformations may redistribute weight among alternatives but may not change the total weight associated with the complete set. This is a separate cross-regime consistency requirement and does not follow from multiplicative record refinement alone. At this stage neither the exponent \(p\) nor any normalization is assumed.

If the weight associated with an amplitude is

\begin{equation}
\mu = |\alpha|^p,
\end{equation}

then the total weight assigned to a complete set of mutually exclusive
records is

\begin{equation}
\sum_{i=1}^{n} |\alpha_i|^p.
\end{equation}

The reversible-invariance requirement is therefore

\begin{equation}
\sum_{i=1}^{n}|(U\boldsymbol{\alpha})_i|^p
=
\sum_{i=1}^{n}|\alpha_i|^p
\end{equation}

for every amplitude vector $\boldsymbol{\alpha}$ and every admissible
reversible transformation $U$.

The relevant finite-dimensional specialization of Lamperti's
isometry theorem is the following \cite{Lamperti1958}. For
$0<p<\infty$, $p\neq2$, an invertible complex-linear transformation
preserving

\begin{equation}
\sum_i |\alpha_i|^p
\end{equation}

for every amplitude vector has the form of a generalized permutation:
it permutes the components and multiplies each component by a complex
factor of unit modulus. Such transformations cannot continuously mix
distinct amplitude components.

For $p=2$, by contrast, the invariant quantity

\begin{equation}
\sum_i |\alpha_i|^2
\end{equation}

is the squared Hermitian norm, whose linear isometry group is the
unitary group. This group contains continuous transformations that
smoothly mix distinct components.

A useful geometric intuition can be given for the special role of
$p=2$. The constant-weight surfaces
$\sum_i|\alpha_i|^2=\mathrm{const.}$ are hyperspheres, which admit
continuous rotations mixing their components while preserving the
total weight. For $p\neq2$, the corresponding linear isometries reduce
to generalized permutations and componentwise phase transformations.

Consequently, $p=2$ is uniquely singled out by compatibility with
continuous reversible linear mixing.

\section{Result}

The preceding analysis first restricts admissible weight assignments
to the family $|\alpha|^p$ and then uses reversible invariance to
select the unique value $p=2$.

Under the stated structural and regularity assumptions, the unique
weight assignment compatible with

\begin{itemize}
\item reversible linear composition,
\item multiplicative representation of record refinement,
\item phase invariance,
\item cross-regime consistency, and
\item continuous reversible linear invariance,
\end{itemize}

is

\begin{equation}
\mu = |\alpha|^2 .
\end{equation}

The uniquely selected weight coincides with the standard quadratic measure of quantum theory. Normalization gives these non-negative weights, with the additive aggregation used above for mutually exclusive alternatives, the mathematical form of a finite probability distribution. The further physical interpretation of this measure as probabilities of individual outcomes or as relative frequencies in repeated experiments depends on the interpretation of quantum measurement and is not assumed in the present derivation.

The derivation shows that the quadratic form arises from the
requirement that outcome weights remain consistently represented
across reversible evolution and irreversible record formation,
without invoking probabilities or expectation values as primitive
quantities.

The emergence of the same quadratic structure from both Hilbert-space
measure theory and compatibility between reversible and irreversible
processes suggests that the Born rule reflects a deeper structural
constraint rather than a feature of any particular formalism.

\section{Conclusion}

The goal of this work was to determine whether the quadratic measure associated with the Born rule can be derived from structural
properties of physical processes rather than introduced as an independent postulate.

We have shown that, under the stated structural and regularity assumptions, the quadratic measure of Born’s rule need not be introduced as an independent probabilistic postulate, but follows from the compatibility between reversible linear evolution and persistent record formation.

The central motivation for this compatibility requirement is that physical processes are not composed exclusively of reversible
evolution, but alternate between reversible dynamics and episodes of record formation. Quantities that remain well defined throughout a
physical process must therefore admit a representation that remains consistent across both regimes. The present derivation leverages the
constraints imposed by this requirement on admissible outcome weights.

The argument takes reversible linear evolution and operationally irreversible record formation as its physical starting point and imposes the structural and regularity conditions stated explicitly in Sec. 2. Weights are associated with possible persistent records and are retained after record formation. Weights are assigned only to
physically realized records, which include the full conditions of their formation. Different interpretations may describe the emergence of such records differently—for example as single-outcome selection or as branching into multiple outcomes—but the underlying structural
distinction between reversible evolution and irreversible record formation remains the same.

A recurring concern in derivations of the Born rule is the possibility of circularity, whereby probabilistic concepts enter implicitly through auxiliary assumptions and subsequently reappear in the final result. The present derivation avoids this route by introducing neither probabilities nor expectation values as primitive quantities. Instead, the argument begins with structural assumptions concerning reversible evolution, record formation, and the compatibility of their respective composition laws. The quadratic measure emerges from these structural requirements rather than from probabilistic postulates.

The present formulation characterizes measurement through the formation of stable records, understood as physically persistent outcomes that remain identifiable after their formation. This shifts the discussion from a concept whose precise role varies across interpretations to a physical process characterized by the formation and persistence of recorded outcomes.

In this way, the Born rule emerges as a structural consequence of the requirement that outcome weights remain consistently represented across reversible evolution and irreversible record formation. The quadratic measure is therefore not introduced as an additional dynamical law, but appears as the unique weight assignment compatible with additive reversible composition, multiplicative record refinement, and continuous reversible invariance.

\section*{Declarations}

\textbf{Author contributions}\\
The author has accepted responsibility for the entire content of this manuscript and approved its submission.

\medskip

\textbf{Use of Large Language Models, AI and Machine Learning Tools	}\\
Chat-GPT was used to improve language, checking consistency, literature search, and translation.

\textbf{Conflict of interest}\\
The author states no conflict of interest

\medskip

\textbf{Research funding}\\
None declared

\medskip

\textbf{Data Availability}\\
Not Applicable


\medskip




\begin{thebibliography}{99}

\bibitem{Deutsch1999}
D. Deutsch, ``Quantum Theory of Probability and Decisions,''
\emph{Proceedings of the Royal Society of London. Series A: Mathematical, Physical and Engineering Sciences},
\textbf{455}, 3129--3137 (1999).
\href{https://doi.org/10.1098/rspa.1999.0443}{doi:10.1098/rspa.1999.0443}.

\bibitem{Zurek2005}
W. H. Zurek, ``Probabilities from Entanglement, Born's Rule $p_k=|\psi_k|^2$ from Envariance,''
\emph{Physical Review A}, \textbf{71}, 052105 (2005).
\href{https://doi.org/10.1103/PhysRevA.71.052105}{doi:10.1103/PhysRevA.71.052105}.

\bibitem{Gleason1957}
A. M. Gleason, ``Measures on the Closed Subspaces of a Hilbert Space,''
\emph{Indiana University Mathematics Journal}, \textbf{6}, 885--893 (1957).
\href{https://doi.org/10.1512/iumj.1957.6.56050}{doi:10.1512/iumj.1957.6.56050}.

\bibitem{Ilyin2016}
A. V. Ilyin, ``The Born Rule and Time-Reversal Symmetry of Quantum Equations of Motion,''
\emph{Foundations of Physics}, \textbf{46}, 845--851 (2016).
\href{https://doi.org/10.1007/s10701-016-0006-y}{doi:10.1007/s10701-016-0006-y}.

\bibitem{Vaidman2020}
L. Vaidman, ``Derivations of the Born Rule,'' in M. Hemmo and O. Shenker (eds.),
\emph{Quantum, Probability, Logic}, Springer, pp. 567--584 (2020).

\bibitem{Neumaier2025}
A. Neumaier, ``The Born Rule---100 Years Ago and Today,''
\emph{Entropy}, \textbf{27}, 415 (2025).
\href{https://doi.org/10.3390/e27040415}{doi:10.3390/e27040415}.

\bibitem{Zhang2026}
J. X. Zhang, ``Summing to Uncertainty: On the Necessity of Additivity in Deriving the Born Rule,''
\emph{Physical Review A} (2026).
\href{https://doi.org/10.1103/mqy7-s2db}{doi:10.1103/mqy7-s2db}.

\bibitem{Horodecki2025}
R. Horodecki, ``Does Quantum Information Require Additional Structure?''
\href{https://arxiv.org/abs/2408.11183}{arXiv:2408.11183} (2025).

\bibitem{Schlosshauer2005}
M. Schlosshauer, ``Decoherence, the Measurement Problem, and Interpretations of Quantum Mechanics,''
\emph{Reviews of Modern Physics}, \textbf{76}, 1267--1305 (2005).
\href{https://doi.org/10.1103/RevModPhys.76.1267}{doi:10.1103/RevModPhys.76.1267}.

\bibitem{Zurek2003}
W. H. Zurek, ``Decoherence, Einselection, and the Quantum Origins of the Classical,''
\emph{Reviews of Modern Physics}, \textbf{75}, 715--775 (2003).
\href{https://doi.org/10.1103/RevModPhys.75.715}{doi:10.1103/RevModPhys.75.715}.

\bibitem{Aczel1948}
J. Acz\'el, ``Sur les op\'erations d\'efinies pour nombres r\'eels,''
\emph{Bulletin de la Soci\'et\'e Math\'ematique de France}, \textbf{76}, 59--64 (1948).
\href{https://doi.org/10.24033/bsmf.1396}{doi:10.24033/bsmf.1396}.

\bibitem{Lamperti1958}
J. Lamperti, ``On the Isometries of Certain Function-Spaces,''
\emph{Pacific Journal of Mathematics}, \textbf{8}, 459--466 (1958).
\href{https://doi.org/10.2140/pjm.1958.8.459}{doi:10.2140/pjm.1958.8.459}.

\end{thebibliography}
\end{document}